\documentclass[11pt]{article}
\usepackage{xspace}
\usepackage{graphicx}
\usepackage{amsmath}
\usepackage{amssymb}
\usepackage{color}

\definecolor{Red}{rgb}{1,0,0}
\definecolor{Green}{rgb}{0,1,0}
\definecolor{Blue}{rgb}{0,0,1}
\definecolor{Black}{rgb}{0,0,0}

\def\beq{\begin{equation}}
\def\eeq#1{\label{#1}\end{equation}}
\def\eeqn{\end{equation}}

\def\beqa{\begin{eqnarray}}
\def\eeqa#1{\label{#1}\end{eqnarray}}
\def\eeqan{\end{eqnarray}}

\let\bar=\overbar

\def\Dslash{\not{\hbox{\kern-4pt $D$}}}
\def\dslash{\not{\hbox{\kern-2pt $\del$}}}

\def\msb{{\bar{\ssstyle M \kern -1pt S}}}

\def\Title#1{\begin{center} {\Large {\bf #1} } \end{center}}

\begin{document}

\Title{UK low-background infrastructure for delivering SuperNEMO}

\bigskip\bigskip


\begin{raggedright}  

{\it Xin Ran Liu\index{Liu, X. R.},\\
Department of Physics and Astronomy\\
University College London\\
WC1E 6BT London, UK}\\

\end{raggedright}
\vspace{1.cm}

\section{Introduction}

SuperNEMO is a next generation neutrinoless double beta decay experiment with a design capability to reach a half-life sensitivity of $10^{26}$ years. This corresponds to an effective Majorana neutrino mass of $\langle m_{\beta\beta} \rangle$ $<$ 50 - 100 meV. 

To achieve the required sensitivity, stringent radio-purity requirements are imposed for both the  construction material and the gas in the tracking chamber. A stringent programme for these materials is therefore required. Dedicated facilities have been established in the UK for screening and selection of detector construction materials. 

\section{SuperNEMO radiopurity strategy for Demonstrator}

SuperNEMO consists of 20 identical modules including a demonstrator module, currently under construction, which contains 7 kg of $^{82}$Se (other isotopes possible). The radiopurity requirements of the demonstrator are A($^{214}$Bi) $<$ 10 $\mu$Bq/kg, A($^{208}$Tl) $<$ 2 $\mu$Bq/kg for the source foil and $<$ 0.15 mBq/m$^3$ for the tracker gas.

In order to achieve these challenging targets all construction materials are screened using High-Purity Germanium (HPGe) detectors to varying levels of sensitivity, depending on location within the detector. Those materials in direct contact with the tracker gas are further screened for radon emanation. Specialised detectors were constructed in order to screen to the sensitivities required by the demonstrator module.

\subsection{Calorimeter}

The demonstrator calorimeter consists of 712 PMTs and scintillator blocks. Materials inside the tracker are typically screened to levels of 0.1 - 10 mBq/kg. Those materials outside the gas volume are less critical, typical sensitivities of 1 - 10 mBq/kg are required for these components.

\subsection{Tracker}

Radioputiry of the tracker gas is essential as radon is one of the most critical backgrounds for SuperNEMO, and most other low background experiments. Radon can enter the detector either through diffusion, contamination during construction or emanation from the detector materials resulting in radioactive daughter isotopes. To reach the target sensitivity the $^{222}$Rn concentration inside the SuperNEMO tracker volume must be less than 150 $\mu$Bq/m$^3$. 

A "Radon Concentration Line" (RnCL) was developed to be used in conjunction with a state-of-the-art radon detector to allow for a more sensitive measurement of large gas volumes. This apparatus has now been commissioned and is capable of measuring radon levels in large samples down to 10 μBq/m$^3$. The results from the first measurements of radon content (gas bottles, boil-off nitrogen and a SuperNEMO sub-module during early stages of construction, etc) using the RnCL are presented. 

\subsection{Source Foil}

Typical source foil will contain 5 kg of double beta decay isotope. The primary candidate currently is $^{82}$Se, with other options such as $^{150}$Nd
and $^{48}$Ca also being considered.

The source foil materials and source frame components are screened using HPGe detectors. A separate dedicated detector, BiPo~\cite{bipo}, was constructed to reach the target sensitivity for the source foil: 2 and 10 $\mu$Bq/kg for $^{208}$Tl and $^{214}$Bi respectively.

\section{Gamma-ray Spectroscopy Facility}

A new low background germanium facility has been established and developed at the Boulby Underground Laboratory. The laboratory is built within a working potash and rock salt mine, the deepest in the UK at 1070 metres (2805 m w.e.)~\cite{robinson}. Due to the low radioactivity in the surrounding salt tunnel the activities of uranium, thorium and radon inside the facility are naturally very low.

There are currently two germanium detectors. The first is an Ortec coaxial ultra-low background high purity germanium detector (HPGe) which has undergone a full refurbishment. A new low background carbon fibre endcap was fitted with a reduced thickness to increase detection efficiency of low energy gammas. The final detector orientation has been changed from the straight to a J configuration. This reduces the number of gammas with a direct line of sight to the germanium crystal, hence minimising the detector background.

The second detector is a new Canberra ultra-low background broad energy germanium detector (BEGe) which replaced a test model of a similar detector. This detector has been designed and optimised to detect low energy gammas with high efficiency. 

\begin{figure}[!ht]
\begin{center}
\includegraphics[width=1\columnwidth]{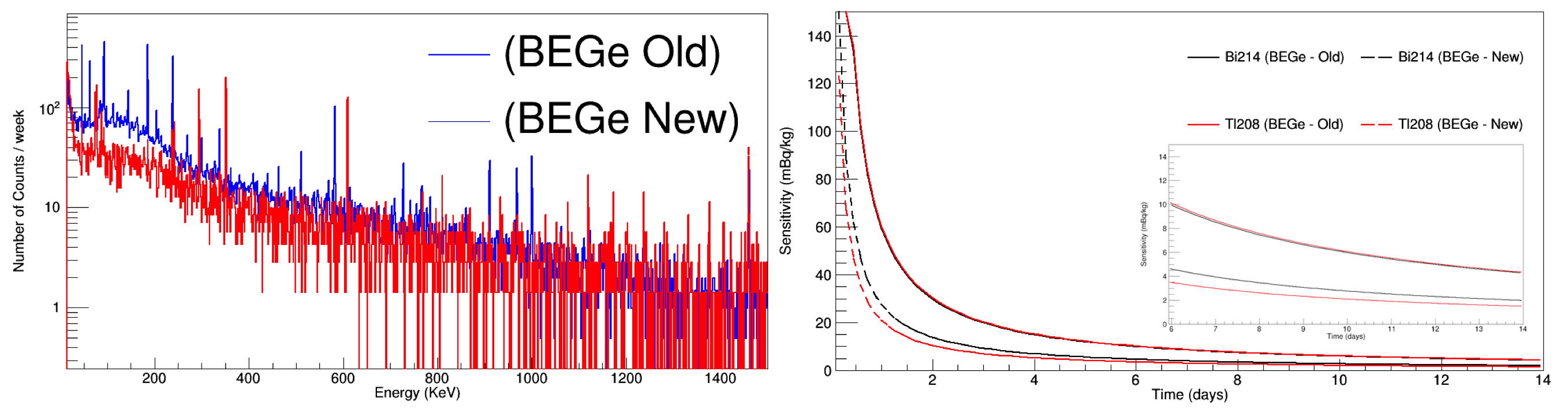}
\caption{Comparison of old and new Canberra BEGe background (left) and sensitivity (right) spectrum.}
\label{fig:detectors}
\end{center}
\end{figure}

Both detectors have been installed within customised shielding with a nitrogen purging system for the central detection cavity. A direct comparison of background spectrums of the new and old BEGe as well as their achievable sensitivities is shown in Figure~\ref{fig:detectors} and demonstrates significant improvement. 

\section{Radon Emanation Facility}

Radon is a critical background for SuperNEMO but samples such as PMTs are not always convenient for measurement using Ge detectors. Emanation chambers are capable of measuring large sample volumes which can compliment germanium spectroscopy and in many cases provide better sensitivity measurements.

A small stainless steel emanation chamber was constructed in Alabama and then assembled and tested at University College London (UCL). The chamber has two flanges, one on each side, which were sealed using copper gaskets, and were tested to ensure no leaks were detected. 

\begin{figure}[!ht]
\begin{center}
\includegraphics[width=1\columnwidth]{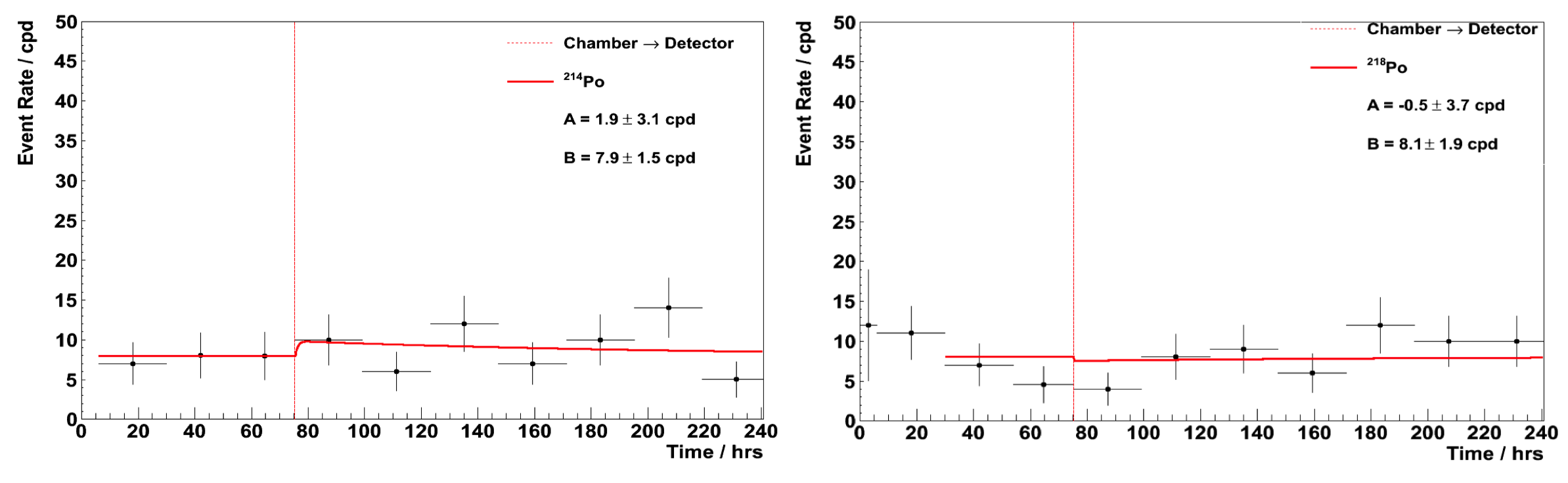}
\caption{Radon emanation chamber background measurement results for $^{214}$Po (left) and $^{218}$Po (right).}
\label{fig:chamber}
\end{center}
\end{figure}

Having established the chamber was well sealed, it was cleaned and resealed to measure the background radon contribution emanating from the chamber itself. The result from the background measurement are shown in Figure~\ref{fig:chamber}.

This chamber, when used in conjunction with a state-of-art electrostatic radon detector~\cite{liu_mott}, results in a sensitivity of $<$ 90 $\mu$Bq and $<$ 120 $\mu$Bq for $^{214}$Po and $^{218}$Po respectively.

\section{Radon Concentration Line}

To reach the sensitivities required for SuperNEMO, $<$ 0.15 mBq/m$^3$, a Radon Concentration Line (RnCL) has been developed. This allows for the monitoring of radon levels within detector sub-modules during construction.

The concept behind the RnCL is to ``concentrate'' radon by pumping a large volume of gas through a cold ultra-pure activated-carbon trap where the radon is adsorbed. The trap is then heated and this concentrated radon is transferred into an electrostatic detector via a helium purge. Using this method, sensitivities of ~0.01 mBq/m$^3$ can be achieved, a 2 orders of magnitude improvement on typical state-of-art stand alone electrostatic detectors.

\subsection{New Radon Trap System}

During radon measurements of large gas volumes one of the greatest source of systematic uncertainty is as a result of the relatively high content of radon in N$_2$, which can be very variable. Previous measurements of cylindered helium and nitrogen have shown variations of up to 1 order of magnitude difference between each batch of gas bottle. 

A radon purification system (RPS) was developed and installed prior to the RnCL in order to remove this source of uncertainty. This was estimated to suppress radon by a factor 1 to 10 orders of magnitude, depending on the carrier gas. This was confirmed by measuring the radon content of cylindered nitrogen after passing through the filter which showed a factor $>$ 20 improvement as shown in Table~\ref{tab:radon}. 

\begin{table}[!th]
\begin{center}
\caption{Result from measurement of radon present in helium and nitrogen gas.}
\begin{tabular}{l|ccc}  \hline\hline
Gas &  Source &  Radon Level  \\ \hline
He  &   Cylinder     &     70-100  \\
N$_2$  &   Cylinder     &     400-1000  \\
N$_2$  &   Boil-off     &     90-140  \\
N$_2$ &  Cylinder +  RPS    &     20  \\ \hline\hline
\end{tabular}
\label{tab:radon}
\end{center}
\end{table}

\section{Summary}

The next generation of rare event search experiments, such as SuperNEMO, will require ultra low activity material screening. The UK infrastructure and screening capacity is being re-developed to provide world-class sensitivity and unique capabilities. These include; establishing a leading low background Ge facility at Boulby Underground Laboratory for material screening and conducting world leading Ge measurements, further developments to increase the radon emanation measurement capacity with further improved sensitivity and continue to push the limit of sensitivity for measuring radon contamination in large gas volumes. 

\bigskip
\section{Acknowledgments}

The author would like to thank Emma Meehan and Sean Paling of Boulby Underground Laboratory for their support and assistance, Craig Theobald and Tom Hunt of the Mullard Space Science Laboratory for their support and James Mott for his pioneering work on the development of the RnCL, and finally to the Rutherford Appleton Laboratory for their funding.

\end{document}